\documentclass[12pt]{article}

\usepackage{amsmath,amsthm,amssymb,amscd}
\usepackage{graphicx, float}
\usepackage[hidelinks]{hyperref}
\usepackage[semicolon,comma]{natbib}
\usepackage{ds}

\usepackage{tabularx} 
\newcolumntype{L}[1]{>{\raggedright\arraybackslash}p{#1}} 

\usepackage{enumitem} 
\usepackage{caption, subcaption} 
\usepackage{capt-of} 
\usepackage[flushleft]{threeparttable} 
\usepackage{multirow}
\usepackage{booktabs}
\usepackage{xcolor,soul} 

\usepackage{authblk}

\title{\sc Economic growth of cities: Does resource allocation matter?}

\author[a]{\sc Sheng Dai}
\author[b]{\sc Timo Kuosmanen}
\author[c,\footnote{Corresponding author. \newline
\hspace*{5mm} \textit{E-mail addresses:} \texttt{sheng.dai@zuel.edu.cn (S. Dai)}, \texttt{timo.kuosmanen@utu.fi (T. Kuosmanen)},\\
\hspace*{34mm} \texttt{zhiqiang.liao@aalto.fi (Z. Liao)}.}]{\sc Zhiqiang Liao}

\affil[a~]{School of Economics, Zhongnan University of Economics and Law, 430073 Wuhan, China}
\affil[b~]{Turku School of Economics, University of Turku, 20500 Turku, Finland}
\affil[c~]{Aalto University School of Business, 02150 Espoo, Finland}
\date{\today}

\begin{document}
\captionsetup[figure]{labelfont={bf},labelformat={default},labelsep=period,name={Fig.}}
\captionsetup[table]{labelfont={bf},labelformat={default},labelsep=period,name={Table}}

\maketitle
 
\vfill
\begin{center}

\end{center}
\vfill

\begin{abstract}
\noindent 
We study how efficient resource reallocation across cities affects potential aggregate growth. Using optimal resource allocation models and data on 284 China's prefecture-level cities in the years 2003--2019, we quantitatively measure the cost of misallocation of resources. We show that average aggregate output gains from reallocating resources across nationwide cities to their efficient use are 1.349- and 1.287-fold in the perfect and imperfect allocation scenarios. We further provide evidence on the effects of administrative division adjustments and local allocation. This suggests that city-level adjustments can yield more aggregate gain and that the output gain from nationwide allocation is likely to be more substantial than that from local allocation. Policy implications are proposed to improve the resource allocation efficiency in China. 
\\[5mm]
\end{abstract}
\vfill

\thispagestyle{empty}

\newpage
\setcounter{page}{1}
\setcounter{footnote}{0}
\pagenumbering{arabic}
\baselineskip 20pt

%

\section{Introduction}\label{sec:intro}

After the reform and opening-up in 1978, China maintained a remarkable average gross domestic product growth rate of 10\% for over three decades. However, following a peak of 13\% in 2007, the country has experienced a downward trend in growth rates \citep{Worldbank2020}. Beyond factors such as trade tensions, high debt levels, and a struggling real estate sector, the declining aggregate productivity growth has been recognized as a key long-term factor that slows down China's economic growth \citep{Worldbank2020, Zhu2024}. This is not alone in China but in virtually all Western countries after the global financial crisis \citep[see, e.g.,][]{Syverson2017, Crafts2018, Dai2023d}. Improving aggregate productivity becomes a central goal for promoting economic growth in both China and other countries alike.

Efficiency gains, a key driver of sustained productivity growth, can be achieved through better resource allocation across production units \citep[see, e.g.,][]{Restuccia2017, Baqaee2019, Dai2023b}. However, nowadays in China, the development of factor markets such as land, labor, capital, and technology lags behind, even though the commodity market has fully realized marketization. Moreover, distortions in factor markets not only exist in the allocation of different types of factors within the same economic entity but also prevail in the allocation between different economic entities \citep[see, e.g.,][]{Hsieh2009, Liu2023, Huang2017}. These persistent misallocations can hinder efficiency gains and, consequently, impede productivity growth. Naturally, it is essential to evaluate the cost of resource misallocation quantitatively.

In a market economy, resources are thought to be efficiently allocated through market competition \citep{Restuccia2017}. However, there can be monopoly power or barriers to free entry as well as government interventions that cause misallocation in a market economy. In a centrally planned economy, the government has greater control over the allocation of resources, but this requires an enormous bureaucracy, and there can be competing interests at the local, regional, and national levels. Further, private investors can influence capital allocation (including foreign firms), and the labor force might migrate illegally. Our paper contributes to a broader debate regarding to what extent the government gets involved with the allocation or just lets the market competition allocate resources. 

What accounts for economic growth differences across countries has been a major concern in macroeconomics, in particular from a resource misallocation perspective \citep[see a review in][]{Restuccia2017}. Yet, the differences across cities or regions within a country, which are equally striking, have been comparatively less focused in the literature. In fact, the economic growth of a country generally relies on the growth of cities \citep{Hsieh2015}. Meanwhile, a wealth of urban economics literature attributes the difference in economic growth across cities to factors such as city size, industrial diversity, human capital, and infrastructure development \citep[see, e.g.,][]{Frick2018, David2013}. Our paper offers an alternative: less-developed cities are not as effective in distributing their production resources to their most efficient use, resulting in significant growth differences across cities. The main finding of this paper confirms that efficient resource allocation across cities can promote aggregate economic growth and narrow regional growth differences.

Our paper relates to a growing literature using causal inference methods to investigate the impact of the administrative division adjustments at China's city or county level on productivity and economic growth \citep[see, e.g.,][]{Bo2020, Deng2022, Han2024}. However, those studies merely demonstrate whether the adjustments can improve economic performance rather than providing a direct quantitative analysis of economic gains from the adjustments. Our paper is also similar to macroeconomic studies using micro data to quantify China's macro development \citep[see, e.g.,][]{Hsieh2009, Brandt2013, Adamopoulos2022}. However, in this paper, we extend it to apply macro data to assess potential economic gains. Considering the role of cities in the growth of a country, we use data on 284 China's prefecture-level cities from 2003 to 2019 and optimal resource allocation models to quantitatively study how administrative division adjustments at the city level affect economic growth.

We also explore whether nationwide resource allocation can bring more economic gains than local or regional allocation. This is relevant because cities at different administrative levels or sizes have different capacities to attract production resources from others. For instance, Shanghai, one of four municipalities in China, can import resources from all other Chinese cities or even other countries, whereas small-sized cities may only attract resources from their neighboring cities. In such a case, redistributing resources at the national level may be more efficient for the entire economy than that at the local or regional level. We further consider the cases that allocate resources locally and show that nationwide resource reallocation yields more aggregate output.

The cost of misallocation measured by potential economic gain is identified to be sensitive to the observation with extreme inputs or outputs \citep{Chen2023}. However, the accuracy of the measurement in such literature as \cite{Hsieh2009}, \cite{Brandt2013}, \cite{Adamopoulos2022}, and \cite{Chen2022} are highly dependent on the single assumed parametric production function, which is known to be sensitive to model misspecification and cannot well capture the heterogeneity of observations \citep{Haltiwanger2018, Dai2023b}. To ameliorate this problem, our paper resorts to multiple nonparametric quantile production functions that inherit fascinating features (e.g., more robust to extremes or outliers) from quantile regression and that consider the heterogeneity explicitly and better predict the changes in the production set \citep{Kuosmanen2020b, Dai2023}.

The rest of the paper is organized as follows. The next section describes optimal resource allocation models. Section~\ref{sec:data} presents the empirical macro data. Section~\ref{sec:res} measures the cost of misallocation and provides a robustness analysis. Further discussion on the impacts of administrative division adjustments at the city level and local allocation is demonstrated in Section~\ref{sec:dis}. We conclude this paper with relevant policy implications in Section~\ref{sec:concl}. 

%

\section{The model}\label{sec:model}

In this section, we set out a theoretical framework to evaluate aggregate output gain in allocating production resources across China's cities. We first apply the convex quantile regression approach to estimate the quantile marginal product of resources at the city level and then design optimal resource allocation models relevant to city development practices and programs.

\subsection{Quantile production functions}\label{sec:qpf}

Consider the prefecture-level cities indexed by $i=1, 2, \ldots, N$ and the yearly time periods by $t=2003, 2004, \ldots, 2019$. In each city, a single good is produced using the physical capital $K_{i,t}$ and labor $L_{i, t}$. Specifically, we assume that the production of $y_{i,t}$ units of good such as gross regional domestic product (GRDP) in city $i$ at year $t$ is based on
\begin{equation}
\label{eq:prod}
    y_{i, t} = f(K_{i, t}, L_{i, t}) + \varepsilon_{i, t}
\end{equation}
where $f$ is an unknown production function to be estimated and $\varepsilon_{i, t}$ is the error term with zero mean and a constant and finite variance. In the context of nonparametric estimation, we do not predetermine the specification of $f$ but assume it to be a continuous, concave, monotonic function. In this paper, our main interest is to estimate the first derivative of $f$ with respect to $K$ or $L$ (i.e., marginal products of resources) and then construct resource allocation models.

In the conventional approach, a single production function is typically used to fit all observations. Consequently, the estimated marginal products may not accurately capture the heterogeneity of observations. We thus resort to convex quantile regression that inherits the robustness to noise and heteroscedasticity from quantile regression \citep{Kuosmanen2020b, Dai2023}. That is, we estimate multiple different quantile production functions and obtain the corresponding marginal products. In so doing, the production function $f$ \eqref{eq:prod} is transformed to the following quantile production production,
$$Q_{y_{i,t}} [\tau \mid K_{i, t}, L_{i, t}] = f(K_{i, t}, L_{i, t}) + F^{-1}_{\varepsilon_{i,t}}(\tau),$$
where $\tau \in (0, 1)$ denotes the given quantile, and $F^{-1}$ represents the inverse cumulative distribution function of the error term. We can then evaluate the marginal products of capital and labor at the relative performance level $ 100\tau\%$ using
\begin{alignat*}{2}
	& \partial Q_{y_{i,t}} [\tau \mid K_{i, t}, L_{i, t}] / \partial K_{i,t}, \\
    & \partial Q_{y_{i,t}} [\tau \mid K_{i, t}, L_{i, t}] / \partial L_{i,t}.
\end{alignat*}

To estimate the yearly quantile marginal products of capital and labor, we solve the following convex quantile regression approach \citep{Kuosmanen2020b} with the given data $\{(L_{i,t}, K_{i,t}, y_{i,t})\}_{i=1}^N$ and quantile $\tau$.

\begin{subequations}
    \begin{alignat}{2}
	   \underset{\alpha,\beta^L, \beta^K, \varepsilon^{+},\varepsilon^{-}}{\mathop{\min }}&\, \tau \sum\limits_{i=1}^{N}{\varepsilon _{i}^{+}}+(1-\tau )\sum\limits_{i=1}^{N}{\varepsilon _{i}^{-}} &{}& \label{eq:cqr_a}\\ 
	   \mbox{subject to} \quad 
	   & y_i=\alpha_i + \beta_i^{L}L_i + \beta_i^{K}K_i +\varepsilon_i^+ -\varepsilon_i^- &\quad& \forall i  \label{eq:cqr_b} \\
          & \varepsilon _i^{+}\ge 0,\ \varepsilon_i^{-} \ge 0 &{}& \forall i  \label{eq:cqr_c}  \\
	     & \alpha_i+\beta_i^{L}L_i + \beta_i^{K}K_i \le \alpha_h + \beta_h^{L}L_i + \beta_h^{K}K_i &{}& \forall i,h   \label{eq:cqr_d}\\
	   & \beta_i^{L} \ge 0,\  \beta_i^{K} \ge 0  &{}& \forall i  \label{eq:cqr_e}
    \end{alignat}
\end{subequations}
where the estimated $\hat{\beta}_i^{K, \tau}$ and $\hat{\beta}_i^{L, \tau}$ are the marginal products of capital and labor for the given quantile $\tau$ at city $i$. In practice, we separately estimate ten different quantiles $\tau \in \{0.05, 0.15, \ldots, 0.95\}$ to obtain quantile marginal products and then use the estimated quantiles to fit the middle part of each decile of the performance distribution (i.e., $0\%-10\%$, $10\%-20\%$, $\ldots$, $90\%-100\%$).

In convex quantile regression, the objective function~\eqref{eq:cqr_a} minimizes the asymmetric absolute deviations from the quantile production function. The first constraint~\eqref{eq:cqr_b} can be interpreted as the multivariate linear regression between inputs and output. The second constraint~\eqref{eq:cqr_c} denotes the sign constraints on residuals. The last two constraints, \eqref{eq:cqr_d} and \eqref{eq:cqr_e}, can guarantee the concavity and monotonicity of quantile production functions. If the constant return-to-scale is assumed on quantile production functions, then the additional constraint $\alpha_i = 0$ is required. Note that linear quantile regression is equivalent to minimizing equation \eqref{eq:cqr_a} subject to constraints \eqref{eq:cqr_b} and \eqref{eq:cqr_c}. 

\subsection{Optimal resource allocation}

To characterize the optimal allocation, we typically assume that in year $t$, a social planner allocates the current aggregate resources into city $i$ to maximize the output of the final good. In the context of heterogeneous modeling, such a generalized resource allocation problem can be solved by utilizing the data-driven quantile resource allocation approach \citep{Dai2023b}. However, the current urban development situation in China calls for more concise modeling settings. In this paper, we consider the baseline models to investigate the effects of resource reallocation and then extend them to analyze the potential aggregate output gain from prefecture-level administrative division adjustments and regional reallocation. 

\textit{Whether resources are allocated perfectly?} In contrast to the ideal allocation, where all resources can be perfectly allocated among cities, more commonly seen cases during the reallocation are that a portion of resources is depleted or not utilized and that an additional iceberg cost is required. That is, the current aggregate resource cannot be perfectly allocated or utilized across cities. For example, installing capital in a country during the reallocation requires an additional iceberg cost \citep{Monge2019}. Further, the existing \textit{hukou} system in China may hinder the free flow of labor resources among cities. We thus model the scenario of whether the current aggregate resources can be fully redistributed among cities. 

To establish optimal resource allocation models, as in \citet{Dai2023b}, we need to assume that the aggregate resources will not increase during the reassignment, that the estimated quantiles represent the technology of each decile, and that cities will keep their efficiency unchanged during the resource reallocation.

Let $\BK_{t}$ and $\BL_{t}$ denote the total aggregate capital and labor of all considered Chinese cities in year $t$. We further split the entire sample of cities into ten mutually exclusive subsamples based on their productive efficiency, representing the ten deciles of the performance distribution. Each subsample includes $N^{\prime}$ cities by construction, where $N^{\prime} = \lceil N/10 \rceil$ and $N$ is the sample size. Consider first the case where the aggregate resources are perfectly allocated across all cities. The planner's problem is to maximize
\begin{subequations}
    \label{eq:lp}
    \begin{alignat}{2}
        \underset{y, k, l} {\mathop{\max }}\, \quad & \sum\limits_{\tau=1}^{10}{\sum\limits_{i=1}^{N^\prime}{y_{i, t}^{\tau} }} &{\quad}& \label{eq:opt1_a}\\
        \mbox{subject to} \quad 
        & y_{i, t}^{\tau} \le \hat{\alpha}_{h,t}^{\tau}+ \hat{\beta}_{h,t}^{K, \tau}k_{i, t}^{\tau} + \hat{\beta}_{h,t}^{L, \tau}l_{i, t}^{\tau}, &{\quad}& \forall i, h, \tau \label{eq:opt1_b} \\
        & \sum\limits_{\tau=1}^{10}{\sum\limits_{i=1}^{N^\prime}{k_{i, t}^{\tau} }} \le \BK_{t} \label{eq:opt1_c}\\
        & \sum\limits_{\tau=1}^{10}{\sum\limits_{i=1}^{N^\prime}{l_{i, t}^{\tau} }} \le \BL_{t} \label{eq:opt1_d}
    \end{alignat}
\end{subequations}
where the counterfactual pseudo-cities in each quantile are indexed by $i=1, \ldots, N^{\prime}$, and the decision variables $k$ and $l$ denote the allocated capital and labor for a pseudo-city $i$ at quantile $\tau$. The objective function \eqref{eq:opt1_a} is set to maximize the total output of all pseudo-cities. The first constraint \eqref{eq:opt1_b} imposes the production capacity on each pseudo-city; that is, the output of a city in each quantile cannot exceed the production technology constraint. The constraints \eqref{eq:opt1_c} and \eqref{eq:opt1_d} simply represent the resource constraint, where the allocated capital and labor cannot exceed their total supply $\BK_{t}$ and $\BL_{t}$. 
 
Note that these counterfactual pseudo-cities do not connect to the real cities other than that they operate using the same quantile production function. One does not need to have the city-specific data on resources, and it suffices to know the aggregate resources (i.e., $\BK$ and $\BL$) and the estimated quantile marginal products (i.e., $\hat{\alpha}^{\tau}$, $\hat{\beta}^{K, \tau}$, and $\hat{\beta}^{L, \tau}$).

\textit{Imperfect allocation:} Consider now a special case where the aggregate resources $\BK$ and $\BL$ are not fully allocated across all cities. Moreover, there may be an unseen depletion in labor or an additional iceberg cost in the capital during the resource reassignment. The planner's problem is to maximize equation \eqref{eq:opt1_a} subject to constraint \eqref{eq:opt1_b} and 
\begin{subequations}
    \begin{alignat}{2}
        & \sum\limits_{\tau=1}^{10}{\sum\limits_{i=1}^{N^\prime}{(1+\lambda_{i, t}^{\tau})k_{i, t}^{\tau} }} \le \BK_{t} \label{eq:opt2_a}\\
        & \sum\limits_{\tau=1}^{10}{\sum\limits_{i=1}^{N^\prime}{(1+\kappa_{i, t}^{\tau})l_{i, t}^{\tau} }} \le \BL_{t} \label{eq:opt2_b}
    \end{alignat}
\end{subequations}
where $\lambda \ge 0$ and $\kappa \ge 0$ are the prespecified parameters denoting the iceberg cost and unseen depletion, respectively. Note that when $\lambda = 0$ and $\kappa = 0$, this problem is equivalent to the first case, where aggregate resources can be fully allocated across cities. For the comparison purpose in counterfactual analysis, we in this paper assume $\lambda = 0.05$ and $\kappa = 0.05$ in a conservative manner. 

From both perfect and imperfect allocation perspectives, it remains unclear which factor—capital or labor—contributes more to aggregate output gains. In other words, we do not know which one exhibits a higher degree of misallocation. Furthermore, in practice, the inputs usually can be classified into two categories: dynamic input that cannot be adjusted in the short term (e.g., capital) and variable input that can be adjusted in the short term (e.g., labor and materials). It thus would be interesting to compare the contributions of capital and labor reallocation. Specifically, we consider the following two sub-scenarios.
\begin{itemize}
    \item Reallocating labor only: The social planner's problems will not include the resource constraint on total capital $\BK$. That is, the constraints \eqref{eq:opt1_c} and \eqref{eq:opt2_a} will be excluded from each problem. In contrast to the above perfect and imperfect allocations (i.e., the long-run reallocation), this scenario reflects short-term reallocation, where the social planner can promptly adjust the allocation strategy.
    \item Reallocating capital only: In this case, the constraint on total labor $\BL$ becomes redundant, and the corresponding constraints \eqref{eq:opt1_d} and \eqref{eq:opt2_b} will be excluded from the social planner’s problems.
\end{itemize}

In summary, we have established four baseline optimal resource allocation models to analyze the aggregate output gain. In the baseline scenario, we consider whether the aggregate resources are allocated perfectly and which factor contributes more to the potential aggregate output. 

\subsection{Aggregate output gain}

To evaluate the effect of the optimal resource allocation, we define aggregate output gain in year $t$ as the ratio of efficient to actual aggregate output,
$$\dfrac{Y_t^e}{Y_t} = \frac{\sum_{\tau}{\sum_i{\hat{y}_{i, t}^{\tau}}}}{\sum_i y_{i, t}},$$
where $Y_t^e$ is the efficient aggregate output in year $t$ estimated by optimal resource allocation models, and $Y_t$ is the actual output aggregated from city-level output $y_{i,t}$. This is a widely used measure of the cost of misallocation in literature \citep[see, e.g.,][]{Hsieh2009, Chen2023, Dai2023b}. Generally, the higher the misallocation level, the larger the aggregate output gain. Further, the reciprocal of  ${Y_t^e}/{Y_t}$ can be used to measure allocative efficiency under the optimal allocation.

We, in practice, estimate the quantile marginal products using the convex quantile regression approach with the Python/PyStoNED package \citep{Dai2021b} and the standard solver Mosek (10.1) and then solve the established optimal resource allocation models by Python/Pyomo package and Mosek and Gurobi solvers.

%

\section{Data}\label{sec:data}

We empirically evaluate the potential aggregate growth in optimal resource allocation using yearly panel data of 284 China's prefecture-level cities from 2003 to 2019.\footnote{
    Other few prefecture-level cities, such as Lhasa, Bijie, and Tongren, are excluded due to insufficient data.
}
To estimate the quantile production functions and solve the resource allocation models, the following input and output variables are specified as
\begin{itemize}
    \item Output ($y$): The city-level GRDP and its price index during the period 2003--2019 were collected from the \textit{China City Statistical Yearbook} and \textit{China Regional Economic Statistical Yearbook}, respectively. The few missing values of the price index were completed by the city-level \textit{Statistical Bulletin of National Economic and Social Development}. We then deflated the nominal GRDP to 2003 constant prices. 
    \item Physical capital ($K$): The classic perpetual inventory method is applied to estimate the city-level physical capital stock. In particular, we resorted to the following formulation  $K_{it} = K_{i,t-1}(1 - \delta_t) + I_{it}/P_{it}$, where $K_{it}$ and $K_{i,t-1}$ represent the capital stocks for city $i$ in year $t$ and $t-1$, respectively. $I_{it}$ and $P_{it}$ denote the total fixed asset investment and its price index for city $i$ in year $t$, respectively. $\delta_t$ is the depreciation rate in year $t$, which is assumed to be 10.96\% \citep{Shan2008}. The capital stock for the initial year is calculated according to $K_{i, 2003}=I_{i, 2004}/(\bar{g}_i + \delta_t)$, where $\bar{g}_i$ is the average growth rate of fixed asset investment for the city $i$ from 2004 to 2019. The \textit{China City Statistical Yearbook} directly provides data on fixed asset investment for 2003--2016, whereas such data for 2017--2019 are unavailable. For data comparability, we first collected the growth rates of fixed asset investment for 2017--2019 from the \textit{Statistical Bulletin of National Economic and Social Development} of each city and then approximated them with 2016 as the base year. Further, the provincial price index $P_{it}$ was applied and collected from the \textit{China Statistical Yearbook}.
    \item Labor ($L$): The data on the number of employees at the end of a year for each city during the period 2003--2019 were obtained from the \textit{China City Statistical Yearbook}.
\end{itemize}

\begin{figure}[t]
    \centering
    \includegraphics[width=\linewidth]{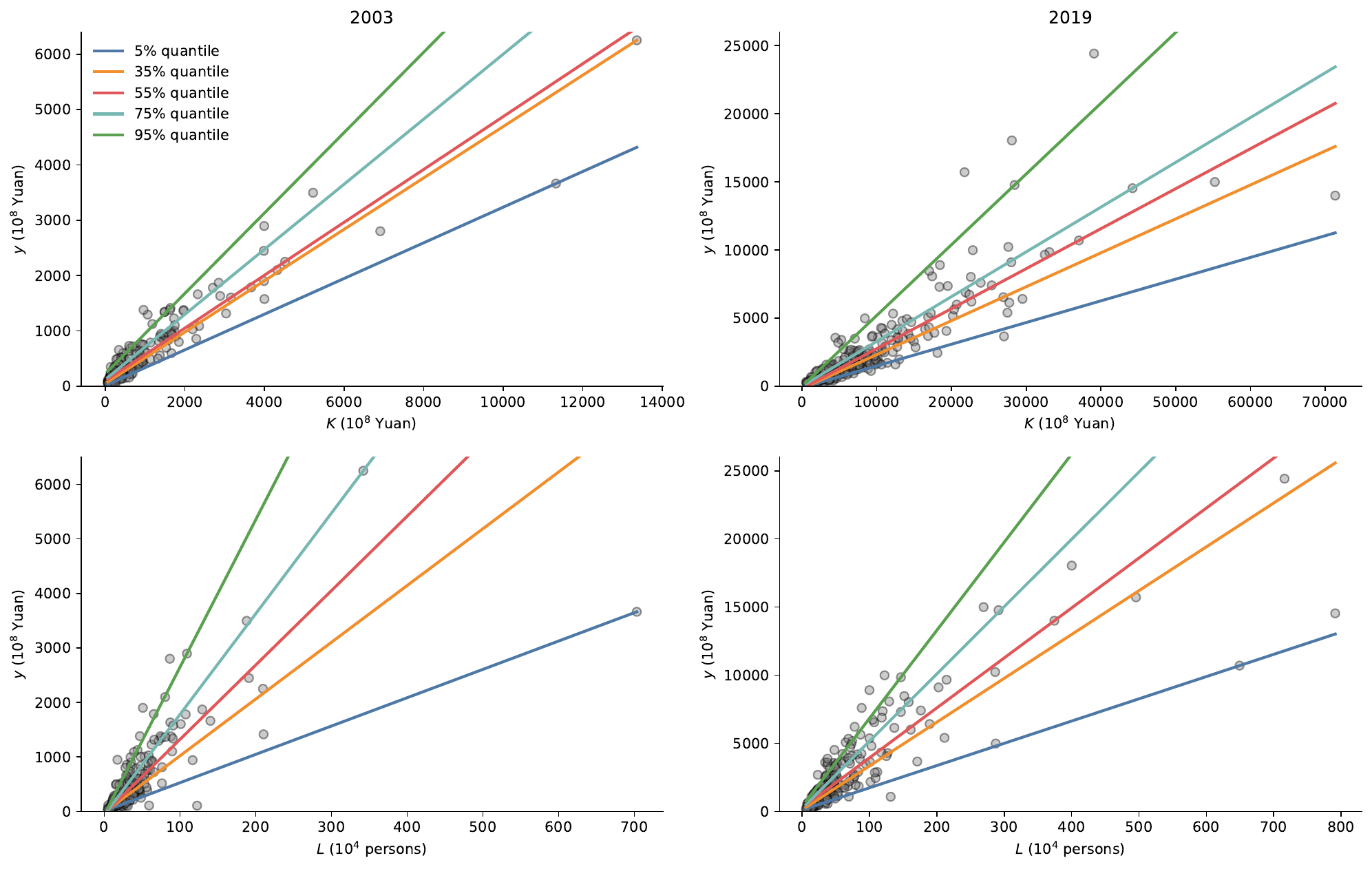}
    \caption{Scatterplots and fitted quantiles of $y$ on $K$($L$).}
    \label{fig:fig1}
\end{figure}

The scatter plots in Fig.~\ref{fig:fig1} show the existence of discrete and extreme observations in the sample. It is also worth noting that the city-level production data deviate from the normal distribution reflected by the estimated kurtosis and skewness statistics (i.e., all values are greater than 0). This implies the presence of high noise and heteroscedasticity in the production data, suggesting that traditional single production function based methods may not be suitable. However, the fitted quantiles of $y$ on $K$ ($L$) in Fig.~\ref{fig:fig1} clearly illustrate how the quantiles help to capture the heterogeneity in capital productivity and labor productivity.

%

\section{Results}\label{sec:res}

We quantitatively measure the cost of resource misallocation across China's cities by combining nonparametric quantile production functions and optimal resource allocation models. We then evaluate the robustness of our main findings with respect to alternative capital stocks, model specifications, and estimation methods. 

\subsection{Aggregate output gain}

\begin{figure}[H]
    \centering
    \begin{subfigure}{\textwidth}
        \centering
        \includegraphics[width=0.8\linewidth]{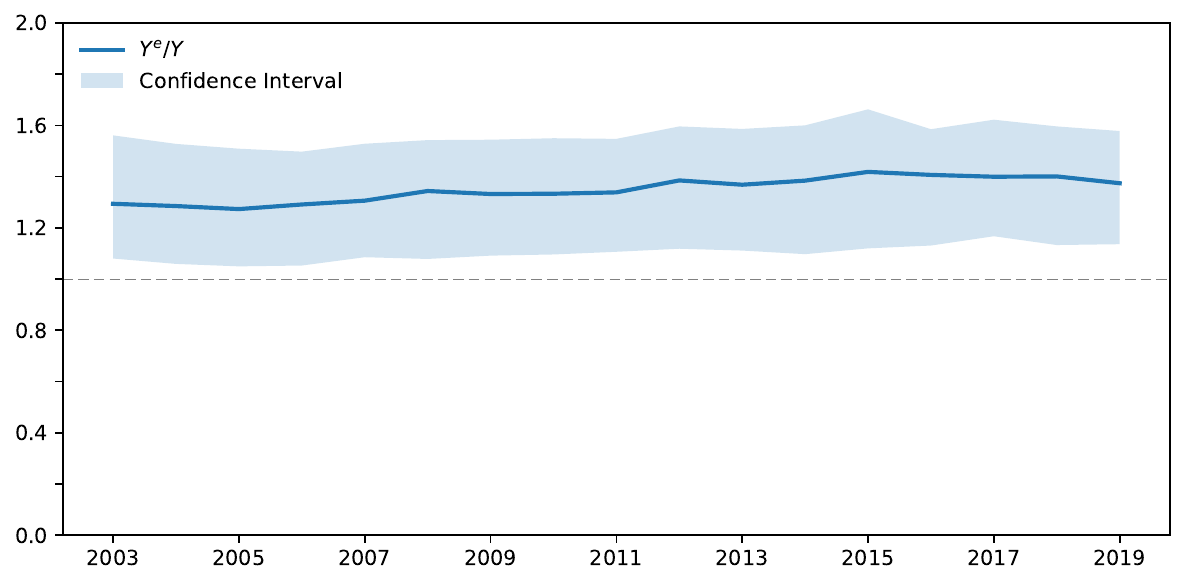}
        \caption{Perfect allocation scenario}
        \label{fig:m1}
    \end{subfigure}
    \hfill
    \begin{subfigure}{\textwidth}
        \centering
        \includegraphics[width=0.8\linewidth]{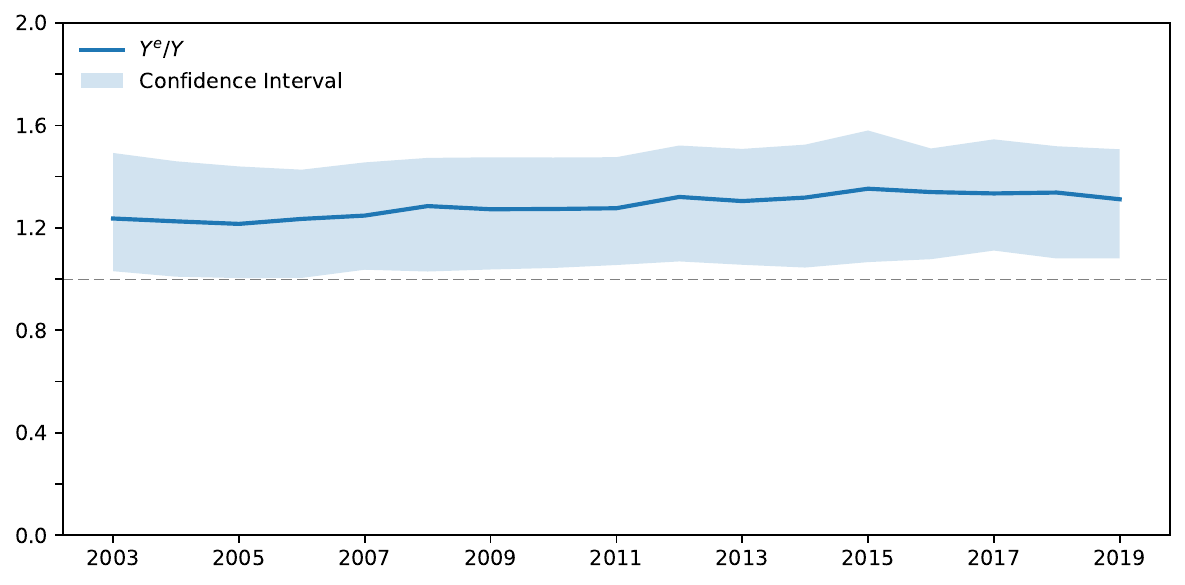}
        \caption{Imperfect allocation scenario (subject to constraints \eqref{eq:opt2_a} and \eqref{eq:opt2_b})}
        \label{fig:m2}
    \end{subfigure}
    \caption{Potential aggregate output growth ($Y^e/Y$).}
    \label{fig:fig2}
\end{figure}

We illustrate the extent of aggregate output gains and their confidence intervals in Fig.~\ref{fig:fig2},\footnote{
    The confidence intervals and standard errors reported in this paper are calculated using bootstrap with 1000 repetitions drawn with replacement at the city level.
} 
where the aggregate resources are reallocated across cities perfectly and imperfectly. The blue line denotes the ratio of the efficient to actual aggregate output ($Y^e/Y$), and the light blue fill indicates the 95\% confidence interval. As shown in Fig.~\ref{fig:fig2}, the output gains in both two scenarios are significantly larger than one in all considered years, indicating that there is resource misallocation at the Chinese city level and that the reallocation can yield more output even in the case where resources are not allocated perfectly. That is, optimal resource allocation can facilitate urban economic growth. We thus would suggest that there is room for productivity improvement by optimal resource reallocation rather than just letting the market competition take care of the allocation.

More specifically, allocating aggregate resources perfectly can result in more output gain. It is evident from Fig.~\ref{fig:fig2} that the average output gain during 2003--2019 is 1.349-fold if the resources are allocated perfectly across cities (i.e., perfect allocation, Fig.~\ref{fig:m1}) and is 1.287-fold if there is a depletion in labor or the additional iceberg cost in the capital during the reallocation (i.e., imperfect allocation, Fig.~\ref{fig:m2}). Further, the yearly output gain in Fig.~\ref{fig:m1} is also larger than that in Fig.~\ref{fig:m2}. In contrast to the ideal perfect allocation, the imperfect allocation seems to be more commonly seen in urban development, but we can also improve allocative efficiency by redistributing resources at the city level.

During the period 2003--2019, we observe that the output gains generally demonstrate an earlier increase and a later decrease trend in both cases. Recall that a higher misallocation level will bring more output gain and vice versa. It is thus no surprise to notice an increasing trend in earlier periods due to factor frictions or distortions, which have been found in, e.g., \citet{Hsieh2009, Brandt2013}. The decreasing trend is perhaps related to the substantial efforts of the Chinese government in recent years. The government has proposed the decisive role of the market in resource allocation and extensive associated measures to ensure that factors could flow freely and independently in an orderly manner and that allocation could be efficient and fair.
\begin{figure}[H]
    \centering
    \includegraphics[width=0.8\textwidth]{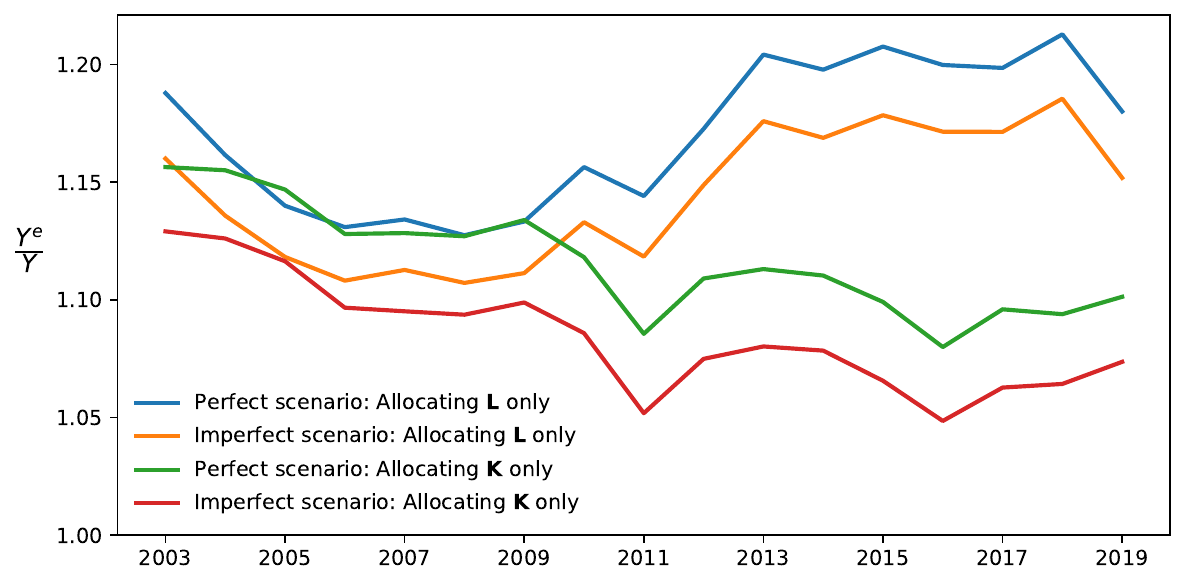}
    \caption{Comparison of the contributions of capital and labor reallocation.}
    \label{fig:fig3}
\end{figure}

We proceed to assess the contributions of capital and labor reallocation to aggregate output gain. Compared to the estimates in Fig.~\ref{fig:fig2}, the resulting output gains in Fig.~\ref{fig:fig3} are relatively smaller but remain above the actual aggregate output when allocating labor or capital merely across cities. It suggests that the optimal long-term allocation always yields a higher $Y^e$ than the optimal short-term allocation. When comparing the contributions of each resource, we observe that labor reallocation generally contributes more to potential aggregate growth than capital allocation, indicating that the misallocation of labor at the city level is more severe than that of capital. This is because greater capital flexibility and more frequent reforms in China's capital market help reduce misallocation. Further, the output gain in perfect resource allocation is literately larger than that in imperfect resource allocation. This confirms the findings in Fig.~\ref{fig:fig2} again.

Considering that the city-level fixed effect for inputs can reduce the input variation, which may be natural or due to potential measurement error \citep{Chen2023}, we further calculate aggregate output gains using the panel data from 2003 to 2019. Table~\ref{tab:tab1} reports the estimates and standard errors under different scenarios. The aggregate output gains from perfect and imperfect allocations are 1.313-fold and 1.252-fold, respectively. Comparatively, these are lower than both yearly and average estimates but still quite substantial in percentage. This is also true for the scenario with the single resource reallocation. Given the fact that the fixed effect estimates of inputs abstract from their transitory variation, the lower output gains of reallocation are potentially the conservative estimates of the cost of misallocation.
\begin{table}[H]
  \centering
  \caption{Aggregate output gain with the fixed effect for inputs.}
    \begin{tabular}{rlrc}
    \toprule
    \multicolumn{2}{c}{\multirow{1}[4]{*}{}} & \multicolumn{1}{c}{\multirow{1}[4]{*}{$Y^e/Y$}} & \multicolumn{1}{c}{\multirow{1}[4]{*}{Standard Error}}  \\
    \multicolumn{2}{c}{}            &       &        \\
    \midrule
    \multicolumn{2}{l}{Perfect allocation}        & 1.313 & 0.12   \\
    \multicolumn{2}{l}{Imperfect allocation}      & 1.252 & 0.11   \\
    \multicolumn{1}{l}{\textit{Allocating $\BL$ only}}              \\
                                   & Perfect allocation   & 1.146 & 0.10   \\
                                   & Imperfect allocation & 1.124 & 0.10   \\
    \multicolumn{1}{l}{\textit{Allocating $\BK$ only}}              \\
                                   & Perfect allocation   & 1.073 & 0.09   \\
                                   & Imperfect allocation & 1.040 & 0.09   \\
    \bottomrule
    \end{tabular}%
  \label{tab:tab1}%
\end{table}%

\subsection{Robustness checks}\label{sec:robust}

In this section, we replace the capital stock, model specification, and estimation method to check whether the main findings in the baseline analysis are robust.

\textit{Alternative capital stock.}---The indirect estimation of capital stock may yield some uncertainty, which mainly originates from the different choices of initial capital stock ($K_{i, 2003}$) or depreciation rate ($\delta_t$). To address this problem, we utilize the alternative initial capital stock and depreciation rate discussed in \citet{Zhang2004} and \citet{Shan2008} to measure capital stock for Chinese cities. Specifically, following \citet{Zhang2004}, the capital stock for city $i$ in the year 2003 is calculated by $K_{i, 2003} = I_{i, 2003}/10\%$, and the depreciation rate is assumed to be 9.6\%, whereas in \citet{Shan2008} the initial capital stock is computed by $K_{i, 2003} = I_{i, 2004}/(\bar{g}_i+\delta_t)$, where $\bar{g}_i$ is the average growth rate of fixed asset investment for city $i$ from 2004 to 2008 and $\delta_t = 10.96\%$. 

With these two alternative capital stock estimates, the resulting output gains are similar to the main findings. The fixed effect results in Table~\ref{tab:tab2} are literally less than both yearly and average results, which confirms that they can be treated as a conservative estimation. Note that the capital stock measured in our paper is slightly larger than that calculated by \citet{Shan2008} but is less than that measured by \citet{Zhang2004}. Compared to results in Table~\ref{tab:tab2} and the previous section, we observe that more capital stock yields less output gain after optimal resource reallocation, implying that the level of capital misallocation is relatively lower. 
\begin{table}[H]
  \centering
  \caption{Robustness check}
  \footnotesize{
    \begin{tabular}{rlrrrrr}
    \toprule
    \multicolumn{2}{c}{\multirow{2}[4]{*}{}} & \multicolumn{2}{c}{Perfect allocation} &       & \multicolumn{2}{c}{Imperfect allocation} \\
\cmidrule{3-4}\cmidrule{6-7}    \multicolumn{2}{c}{} & \multicolumn{1}{l}{Average} & \multicolumn{1}{l}{Fixed effect} &       & \multicolumn{1}{l}{Average} & \multicolumn{1}{l}{Fixed effect} \\
    \midrule
    \multicolumn{1}{l}{Alternative capital stock} & \citet{Zhang2004} &  1.332  &  1.308    &       & 1.271      & 1.247 \\
          & \citet{Shan2008}    &  1.366    &  1.310    &           &   1.302   & 1.247 \\
          &                     &           &           &           &           &  \\
    \multicolumn{1}{l}{Alternative specification}  &  Human capital        &           &           &           &  \\
          & \hspace*{1em} \textit{A separate input}    &   1.371   &   1.336   &       & 1.302      & 1.260       \\
          & \hspace*{1em} \textit{Together with labor} &    1.317  &  1.273    &           &   1.259   &  1.203 \\
          & Natural resource    &         &        &        &        &  \\
          & \hspace*{1em} \textit{Land input}  &  1.353     &  1.329    &        & 1.289    & 1.262 \\
          &                     &           &           &           &           &  \\
    \multicolumn{1}{l}{Alternative method} & \citet{Chen2023}       &   1.067   &  1.049    &      & ---     & --- \\
    \bottomrule
    \end{tabular}%
    }
  \label{tab:tab2}%
\end{table}%

\textit{Alternative model specification.}---Given the facts that human capital can enhance the efficiency of resource allocation and that natural resources remain a substantial aspect of production in less developed cities, we further explore the optimal resource allocation under alternative model specifications involving human capital and natural resources (e.g., land input). Specifically, we use the average years of education to measure the degree of human capital\footnote{
    $\textit{Average Years of Education}=6S_1+10S_2+16S_3$, where $S_1$, $S_2$, and $S_3$ represent the number of students per ten thousand in primary schools, ordinary secondary schools, and higher education institutions, respectively. 
} 
and take the year-end urban construction land area as the land input. We collected the original data from the \textit{China City Statistical Yearbook} (2004--2020). Furthermore, human capital can be regarded as a separate input \citep[e.g.,][]{Dai2023d} or together with labor as the human capital-augmented labor \citep[e.g.,][]{Monge2019}.

When considering human capital as a factor together with labor in baseline models, we have similar results to those in the main findings. However, after introducing human capital or land into the models as an additional input, we obtain a slightly higher output gain, implying that there are misallocations of human capital and land resources. This confirms the relevant findings in, \citealp[e.g.,][]{Huang2017, Chen2021c, Liu2023}. In such cases, more resources can be reallocated and thus yield more outputs.

\textit{Alternative method.}---We also consider the conventional parametric method to quantify the cost of resource misallocation. Following \citet{Chen2023}, we apply the parametric framework to solve social planner problems relying on the first-order conditions. While the magnitudes of output gain are less than our estimates, they also show the resource misallocation across Chinese cities and the positive role of resource reallocation in urban growth. The different output gain estimates may be attributed to the following factors:\footnote{
    In addition, the different calibrated parameters (e.g., the value of the ``span of control'' parameter) could also affect the final estimates.
} 
1) in contrast to the parametric framework that relies on a single production function, our estimates use multiple quantile production functions to take the heterogeneity of cities into account and utilize the full information of each city; 2) The assumption of constant return-to-scale on production function is considered in their parametric framework, whereas we assume production functions to be variable return-to-scale; 3) to quantify the output gain, we solve social planner problems by computing full-scale optimization models rather than the heuristic first-order conditions, which are also likely to be sensitive to the model specification \citep{Osotimehin2019}. Nevertheless, the detailed comparison of first-order condition based and full-scale optimization based estimations warrants further research.

%

\section{Extensions}\label{sec:dis}

We further extend our analysis to discuss the impacts of the city-level administrative division adjustments and local allocation. 

\subsection{Administrative division adjustments}

Since the reform and opening-up, especially since entering the 21\textsuperscript{st} century, with the deepening of the national development strategy and the continuous rapid industrialization and urbanization process, the demand for urban spatial expansion has increased synchronously, and the need and frequency of administrative division adjustments have also been continuously enhanced. During the period 2000--2019, there were 15 substantial splits and mergers in the administrative division adjustments of Chinese prefecture-level cities.\footnote{
    For example, Laiwu (a former prefecture-level city) was adjusted to one of the subdistricts of Jinan (a prefecture-level city) in January 2019.
    } 
Further, there were numerous records of adjustments to the county-level cities due to the incorporating counties into prefectures reform (\textit{chexian shequ}). It is well recognized that such adjustments can facilitate urban economic growth and development performance \citep[see, e.g.,][]{Fan2012, Wang2020e, Han2024}. However, the impact of administrative division adjustments across cities on resource allocation is often left unexplored. We thus discuss the effects of \textit{whether administrative division adjustments occur}.

To model the scenario where a city is allowed to be merged, we reformulate the baseline social planner's problem as
\begin{subequations}
    \label{eq:milp}
   \begin{alignat}{2}
        \underset{\bx, y, b} {\mathop{\max }}\, \quad & \sum\limits_{\tau=1}^{10}{\sum\limits_{i=1}^{n}{y_i^{\tau} }} &{\quad}& \label{eq:MIP}\\
        \mbox{subject to} \quad
        & y_{i, t}^{\tau} \le \hat{\alpha}_{h,t}^{\tau}+ \hat{\beta}_{h,t}^{K, \tau}k_{i, t}^{\tau} + \hat{\beta}_{h,t}^{L, \tau}l_{i, t}^{\tau} + (1-b_i)M &{}& \forall h,i,\tau  \label{eq:opt5_a}\\
        & y_{i, t}^{\tau} \le b_{i, t}M, k_{i, t}^{\tau} \le b_{i, t}M, l_{i, t}^{\tau} \le b_{i, t}M &{}& \forall i,\tau  \label{eq:opt5_b} \\
        & b_{i, t} \in \{0, 1\} \label{eq:opt5_c}\\ 
        & \sum\limits_{\tau=1}^{10}{\sum\limits_{i=1}^{N^\prime}{k_{i, t}^{\tau} }} \le \BK_{t} \label{eq:opt5_d}\\
        & \sum\limits_{\tau=1}^{10}{\sum\limits_{i=1}^{N^\prime}{l_{i, t}^{\tau} }} \le \BL_{t} \label{eq:opt5_e}
\end{alignat} 
\end{subequations}
where $M$ is a prespecified large positive number and determined by the natural
bounds provided by the problem itself \citep{Dai2023b}. Compared to the baseline problem \eqref{eq:lp}, the extended planner's problem \eqref{eq:milp} introduces a set of binary constraints \eqref{eq:opt5_c} to represent the ``entry and exit'' of a city. That is, if $b=0$, then pseudo-city $i$ is allowed to be merged.

The dynamic force of creative destruction (i.e., administrative division adjustments)  is evident in this scenario. When considering the possibility of administrative division adjustments among cities in the planner's problem, we observe the aggregate output gain becomes larger than that without administrative division adjustments. The average aggregate output gain from 2003--2019 is 1.426-fold, and the lowest and largest output gains are observed in 2005 and 2015 (i.e., 1.316-fold and 1.512-fold, respectively). Further, the aggregate output gain in the fixed effect model is 1.398-fold. Those suggest that the administrative division adjustments among Chinese cities can yield more aggregate output. This is similar to the study on the impact of exit and entry of firms, where redistributing production resources between active firms has a positive effect on aggregate productivity or output \citep[see, e.g.,][]{Brandt2012, Jaef2018}.

Several factors may explain the increase in aggregate output following the implementation of administrative division adjustments. First, by adjusting administrative divisions, the Chinese government can more effectively allocate resources (e.g., capital, labor, and infrastructure) to regions with higher growth potential or strategic importance. This optimized resource allocation boosts productivity and output, particularly in larger cities. Second, these adjustments enable cities to specialize in industries where they hold a comparative advantage, fostering economies of scale, enhancing efficiency, and ultimately increasing output. Third, redrawing administrative boundaries can streamline governance, allowing local authorities to better implement policies tailored to their region's specific needs. Improved governance creates a more supportive environment for economic growth and development. Finally, administrative adjustments are often accompanied by infrastructure investments—such as in transportation, utilities, and telecommunications—which reduce transaction costs, facilitate trade, and attract further investment, all contributing to higher aggregate output.

\subsection{Local allocation} 

Each city in China has its unique administrative level or rank based on its political status, economic level, and geographical location. This system has clearly benefited the central or high-level cities \citep{Ma2005, Cheng2019}, which have more potential and capacity to attract nationwide resources. In the process of resource allocation, for example, Shanghai has a large probability of obtaining resources from any city in China, whereas Nanjing may only be assigned resources from cities in the Yangtze River Delta region or neighbor cities. That is, the heterogeneity in urban resource endowments and socioeconomic characteristics engender disparate capacities for resource redistribution among cities. It is then natural to consider how national- or regional-wide allocation affects the optimal allocation. We thus further discuss the impacts of \textit{whether resources are allocated nationally}.

To incorporate this scenario into the planner's problem \eqref{eq:lp}, we impose the following two resource constraints to problems \eqref{eq:lp} or \eqref{eq:milp}.
\begin{subequations}
   \begin{alignat}{2}
    & \sum\limits_{i=1}^{N^\prime}{l_{i, t}^{\tau}} \le \BL_{t}/10 \quad \forall \tau \\
    & \sum\limits_{i=1}^{N^\prime}{k_{i, t}^{\tau}} \le \BK_{t}/10 \quad \forall \tau 
\end{alignat} 
\end{subequations}
where the additional constraints restrict resources to be distributed across cities with similar production technology. In contrast to problems \eqref{eq:lp} or \eqref{eq:milp} where resources are allocated to a nationwide pseudo-city $i$, the new planner's problem can ensure that resources will be reallocated to pseudo-city $i$ from a local region.

\begin{figure}[H]
    \centering
    \includegraphics[width=0.8\textwidth]{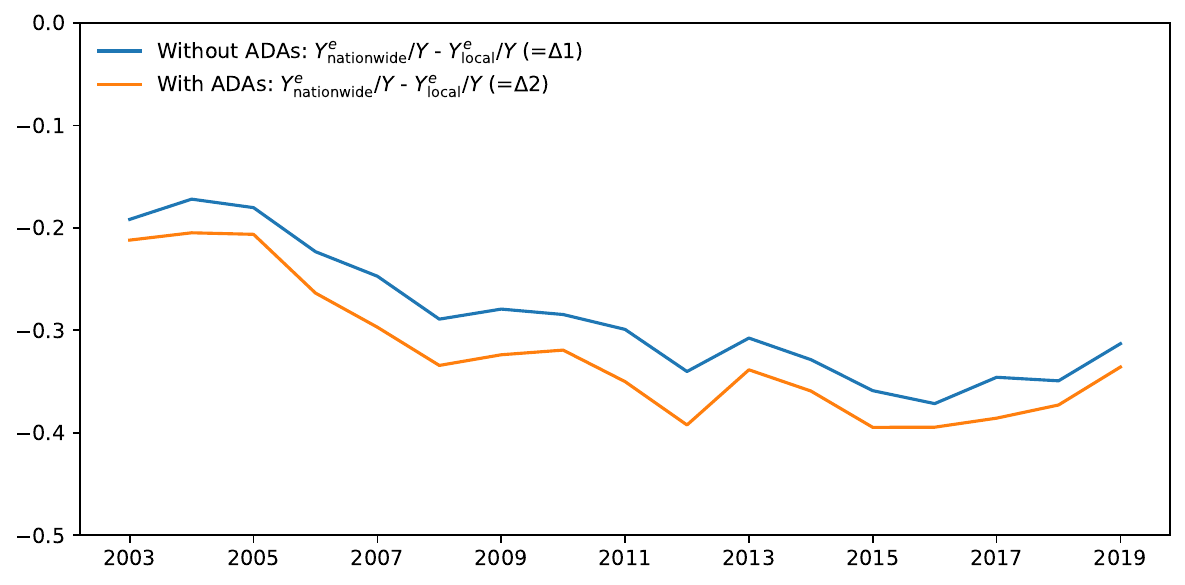}
    \caption{Difference in output gain between nationwide and local allocations with and without administrative division adjustments (ADAs).}
    \label{fig:fig4}
\end{figure}

Fig.~\ref{fig:fig4} demonstrates the difference in aggregate output gain between nationwide and local allocations with and without administrative division adjustments. $\Delta1$ in Fig.~\ref{fig:fig4} denotes the difference in output gain between nationwide and local allocations without administrative division adjustments, and $\Delta2$ represents the difference in output gain between nationwide and local allocations with administrative division adjustments. As shown in Fig.~\ref{fig:fig4}, the values of $\Delta2$ are less than those of $\Delta1$, confirming again that allowing administrative division adjustments can yield more potential output.  

We also observe that the values of both lines in all years are negative, indicating that the aggregate output generated by the optimized allocation of resources nationwide is higher than that by the allocation of resources in local areas. This finding is intriguing and aligns with the expectations of current policies promoting market-oriented reforms in resource allocation. In the early stages, both central and local governments implemented a series of policies aimed at facilitating the flow of resource factors and reducing misallocation in certain cities or local regions, resulting in some noticeable policy benefits. However, as expectations for greater social welfare and economic output grow, nationwide reforms promoting the marketization of resource factors have been further advanced. With the enhancement of marketization levels in resource factors, efficient allocation of resources nationwide becomes possible, and policy benefits are expected to increase significantly.

%

\section{Conclusions}\label{sec:concl}

Misallocation of resources is a widely acknowledged explanation of why some economies are rich and others poor. That is, such differences in economic growth across economies could be offset by the optimal resource allocation. In this paper, we study how efficient resource reallocation affects potential aggregate growth by a counterfactual analysis framework. Using optimal resource allocation models and data on 284 China's prefecture-level cities in the years 2003–2019, we quantitatively measure the cost of misallocation of resources and explore the impacts of efficient resource reallocation across cities, the city-level administrative division adjustments and local allocation. 

Our baseline empirical estimates show that the average aggregate output gains from reallocating resources across nationwide cities to their efficient use are 1.349- and 1.287-fold in the perfect and imperfect allocation scenarios. This suggests that if production resources are allocated across cities efficiently, whether a perfect or imperfect allocation, the aggregate economic gains would far exceed the sum of the current economic output. Therefore, it is highly recommended that central and local governments continuously improve resource allocation efficiency by guiding the flow of capital and labor, strengthening inter-regional cooperation to reduce market segmentation, and maintaining the vitality of market entities.

When the administrative division adjustments at the city level are allowed, the potential economic gain is higher than the case without adjustment. That is, appropriate urban planning adjustments can help increase the overall economic output of the country. This could provide new insights for local governments in future urban development and urban practices. For example, governments can optimize urban spatial layouts to improve efficiency in urban space and resource utilization and reduce congestion in the process of market-oriented resource allocation. Furthermore, it is necessary to gradually improve the efficiency of the administrative management system, streamline administrative approval procedures, and enhance government service efficiency and quality.

The aggregate output gain generated from optimal resource allocation at a national scale far outweighs those from resource allocation solely within regions. Hence, in the reform of market-oriented resource allocation, it is essential to accelerate the establishment of a unified nationwide factor market. Specifically, the central government needs to formulate unified market access standards and rules to eliminate regional differences, lower barriers for enterprises to enter the market, and promote the free flow and optimal allocation of resources. A unified regulatory system and legal framework also need to be established to ensure a fair competitive environment, prevent monopolies and unfair competition, and enhance market transparency and predictability. Furthermore, local governments could simplify cross-border trade procedures and processes, reduce trade costs and time, and facilitate the free flow of goods, capital, and information.

While modeling resource allocation based on centrally planned systems can provide valuable insights into free market dynamics, it may not fully capture the complexities of real-world market scenarios, particularly in situations that deviate from perfect competition. Moreover, although our study reveals significant resource misallocation at the city level, these findings may not be directly applicable to individual cities for their productivity improvement. These important aspects are left for future research.

%

\section*{Acknowledgments}\label{sec:ack}

The authors wish to acknowledge CSC – IT Center for Science, Finland, for computational resources. Sheng Dai gratefully acknowledges financial support from the Foundation for Economic Education (Liikesivistysrahasto) [grant no.~230110] and the OP Group Research Foundation [grant no.~20230008]. Zhiqiang Liao gratefully acknowledges financial support from the Jenny and Antti Wihuri Foundation [grant no. 00220201] and the Foundation for Economic Education (Liikesivistysrahasto) [grant no. 230261].

%

\baselineskip 12pt
\bibliographystyle{econ-econometrica}
\bibliography{References.bib} 

\end{document}